\documentstyle[epsfig,aps,floats,twocolumn]{revtex}

\def\beq{\begin{equation}}
\def\eeq{\end{equation}}

\def\kvec{{\bf k}}
\def\half{{\textstyle{1\over 2}}}

\def\etal{{\it et al}}

\def\shalf{{\scriptstyle{1\over 2}}}
\def\mpi{M_\pi}
\def\mn{m_{\scriptscriptstyle N}}
\def\N{{\scriptscriptstyle N}}
\def\A{{\scriptscriptstyle A}}
\def\Mojzis{Moj\v{z}i\v{s}}
\def\slash#1{#1\llap/}

\begin{document}
\draft

\twocolumn[\hsize\textwidth\columnwidth\hsize\csname@twocolumnfalse\endcsname
\title{On the absence of fifth-order contributions to the nucleon mass\\ in 
heavy-baryon chiral perturbation theory}
\author{Judith A. McGovern
and Michael C. Birse}
\vskip 20pt
\address{Theoretical Physics Group, Department of Physics and Astronomy\\
University of Manchester, Manchester, M13 9PL, U.K.}
\nopagebreak
\maketitle
\begin{abstract}
We have calculated the contribution of order $\mpi^5$ in the chiral expansion of
the nucleon mass in two-flavour heavy-baryon chiral perturbation theory.
Only one irreducible two-loop integral enters, and this vanishes.  All other
corrections in the heavy-baryon limit can be absorbed in the physical
pion-nucleon coupling constant which enters in the $\mpi^3$, term, and
so there are no contributions at $\mpi^5$. Including finite nucleon mass
corrections, the only contribution agrees with the expansion of the relativistic
one-loop graph in powers of $\mpi/\mn$, and is only 0.3\% of the $\mpi^3$ term.
This is an encouraging result for the convergence of two-flavour heavy-baryon
chiral perturbation theory.
\end{abstract}
\pacs{}]

\bigskip

The fundamental degrees of freedom of the strong interaction are quarks and
gluons, but in spite of the many successes of QCD in describing high energy
phenomenology, a full description of the particles that constitute ordinary
matter still eludes us.  There is however increasing interest in the
interactions of such particles at low energies as new, precise, data on nucleon
properties and interactions becomes available.  For low enough energy it has
long been known that the interaction of pions is governed only by the symmetries
of QCD, in particular SU(2)$\times$SU(2) chiral symmetry, and a successful
systematic effective field theory of pions, Chiral Perturbation Theory 
\cite{GL}, has been developed.  In this theory the pionic lagrangian is 
expanded as a power series, with each subsequent term having two more
derivatives or powers of the pion mass than the previous one, and with the
lowest order being two: ${\cal L}_{\pi\pi}={\cal L}_{\pi\pi}^{(2)}
+{\cal L}_{\pi\pi}^{(4)}+\ldots$. The expansion is in powers of $q/\Lambda$,
where $q$ is a momentum or pion mass, and $\Lambda$ is the scale of the physics
which has not been included explicitly, for instance the $\rho$ meson.
Because chiral symmetry requires the
pion-pion scattering length to vanish in the chiral limit as the pion momentum
goes to zero, loop diagrams with vertices from ${\cal L}_{\pi\pi}^{(2)}$
only contribute at fourth order; divergences are cancelled by counterterms in 
${\cal L}_{\pi\pi}^{(4)}$, and so on.  At fourth order eight {\it a priori}
undetermined low energy constants (LEC's) enter, which have to be fit to 
data.  The current state of the art is calculation to sixth order, involving
two-loop integrals \cite{bij}.

Attempts to expand the success of the pionic theory by including nucleons 
initially ran into difficulties; since the nucleon mass is not small, it spoils
the expansion of physical amplitudes in powers of small quantities (masses and
momenta), and the theory is not systematic \cite{GSS}.  However if the nucleon
mass is taken to be infinite, it effectively decouples, and the power counting
is restored. The Lagrangian expansion includes both odd and even numbers of 
derivatives, ${\cal L}_{\pi\N}={\cal L}_{\pi\N}^{(1)}+{\cal L}_{\pi\N}^{(2)}
+\ldots$.  Only even powers of the pion mass enter, since 
the underlying parameter is the quark mass; $\mpi^2\sim M_q$.
Furthermore corrections for a finite mass can be included
systematically, with a simultaneous expansion in powers of $1/\mn$; since $\mn
\sim\Lambda\sim M_\rho$ these terms can naturally be fitted into the expansion 
above; thus ${\cal L}_{\pi\N}^{(2)}$ contains terms of order $1/\mn$, and 
${\cal L}_{\pi\N}^{(3)}$, of order $1/\mn^2$, etc.

The heavy-baryon expansion starts by splitting the nucleon momentum into two
parts, $p_\mu=m v_\mu+l_\mu$, with $v^2=1$ and $v\cdot l\ll m$. The projection
operators $P_v^\pm=(1\pm\slash{v})/2$ are used to split the nucleon spinor into 
two parts, $\Psi=e^{-i m v\cdot x}(H+h)$, and the ``small-component" degrees of
freedom $h$ are integrated out.  For details, the reader should consult, for
instance, Ref.~\cite{mei95}.

Heavy-baryon chiral perturbation theory (HBCPT) has been applied to a number of
problems involving the electromagnetic properties of nucleons and the
interaction of nucleons and pions.  The number of terms to a given order
however is much larger than in the pionic theory; there are for instance
seven LEC's in ${\cal L}_{\pi\N}^{(2)}$ and twenty three in 
${\cal L}_{\pi\N}^{(3)}$.  So far the full Lagrangian has only been worked out
to third order, though individual terms have been considered at higher order.
By no means all of the LEC's to third order have been determined from
experiment. Although many of the calculations which have been done show the
method to be promising, the convergence of the expansion of
amplitudes in powers of the physical pion mass is hard to judge so far.  In
particular, almost all calculations so far have been to one-loop order
only\footnote{Bernard {\it et al.}\ \cite{mei96} consider the
contribution of two-loop diagrams to the imaginary part of the nucleon 
isoscalar electromagnetic formfactor, but this does not require the evaluation
of two-loop integrals.}.

One of the quantities which, through its relative simplicity, lends itself
to such an analysis, is the nucleon mass shift due to the finite quark
(pion) mass, or equivalently the pion-nucleon sigma commutator,
$\sigma_{\pi\N}=\mpi^2\partial \mn/\partial \mpi^2$,  estimated from scattering
data to be $45\pm8$~MeV \cite{GLS}.   To order $q^3$ there are two
contributions, with one LEC which has now been estimated independently from
pion-nucleon scattering data \cite{moj98,mei98}.  To order $q^4$ more LEC's
will enter, for which there we as yet have no experimental handle.  Borasoy and
Mei\ss ner have attempted to estimate fourth order LEC's in three flavour 
HBCPT through the principle of resonance saturation \cite{bor}. There are
however good reasons to distrust three flavour calculations, at least without
an explicit decuplet, since nucleonic excitations which are being kept (such as
$\Sigma K$) are much higher in energy than others which have been integrated
out (the delta and higher resonances) \cite{mil}. (Convergence of the
three-flavour result also appears to be poor\cite{hol}.) However the mass shift 
is one quantity where it is possible to look at the order $q^5$ term without 
knowing the $q^4$ piece, and this is what we have done in this paper.

In order to calculate most quantities to order $q^5$, the expansion of the
nucleonic Lagrangian up to ${\cal L}_{\pi\N}^{(5)}$ would be required. The only
contribution to $\Sigma(0)$ from the fifth order term would be a simple
counterterm of order $\mpi^5$. However
the Lagrangian is analytic in the quark masses, that is in $\mpi^2$, so such a
term cannot exist.  (For the same reason any irreducible two-loop diagrams must
give finite contributions to $\Sigma(0)$, since there can be no counterterm
to cancel divergences.) Similarly, in the absence of mass insertions of order
$\mpi$, ${\cal L}_{\pi\N}^{(4)}$ cannot contribute at this order.  
All the relevant Feynman amplitudes for these calculations can be found
in  the work of Mei\ss ner \etal \cite{mei95,mei98} or, using an
alternative form of ${\cal L}_{\pi\N}^{(3)}$, in that of \Mojzis 
\cite{moj98} (who also gives the relevant amplitudes from  
${\cal L}_{\pi\pi}^{(4)}$).

\begin{figure}
  \begin{center} \mbox{\epsfig{file=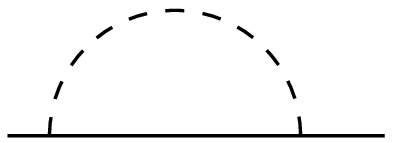,width=4.0truecm,angle=0}}
\end{center}

{\bf Fig.~1:} The one-loop diagram of $\Sigma^{(3)}$.  Solid lines are nucleons and 
dashed lines, pions.
\end{figure}

The heavy-baryon propagator is given by 
\beq 
S^{-1}=\omega-\Sigma(\omega,\kvec),
\eeq
where the nucleon momentum is written as $p=m v+k$, $m$ is the bare mass, and
$\omega=v\cdot k$.  The mass shift $\delta m=\mn-m$ is the value of $\omega$
for which the propagator has a pole at zero three-momentum:
\beq 
\delta m -\Sigma(\delta m,0)=0.
\label{mass_shift}
\eeq
(Of course the mass shift could be found from the pole of the 
propagator for any three-momentum, but as HBCPT is constructed to respect 
Lorentz invariance, the result will not change.)

In order to solve Eq.~\ref{mass_shift} to a given order in $\mpi$, both
$\delta m$ and $\Sigma(\omega)$ must be expanded in powers of $\mpi$.  
To order $\mpi^3$,\ \  $\Sigma(\delta m)=\Sigma(0)$, so \cite{mei95}
\beq
\delta m^{(2)}+\delta m^{(3)}=-4 c_1 \mpi^2 -{3 g_\A^2\mpi^3\over 32\pi 
F_\pi^2 },
\label{mass23}
\eeq
where the second term comes from the diagram in Fig.~1.
Writing
as $\Sigma^{(n)}$ the expression for the $O(q^n)$ part of $\Sigma$, which will
have an expansion  $\Sigma^{(n)}(\omega)= a_1\mpi^n +a_2\mpi^{(n-1)} \omega
+\ldots$, we obtain
\begin{eqnarray}
\delta m^{(5)}=\Sigma^{(5)}(0)&+&\delta m^{(2)}{\Sigma^{(4)}}'(0)+\delta m^{(3)}
{\Sigma^{(3)}}'(0)\nonumber
\\&&+\half (\delta m^{(2)})^2 {\Sigma^{(3)}}''(0)
\label{mass5}
\end{eqnarray}
where primes indicate derivatives with respect to $\omega$.
Any calculation in HBCPT yields an answer in terms of the bare Lagrangian
parameters,  $M, g, F$ and $m$ to lowest order, which are the first terms in an
expansion in powers of $M$ of $\mpi, g_\A, F_\pi$ and $\mn$.  It is
customary to replace the bare parameters by the physical ones so that the
lowest order predictions do not change as higher orders are added.  This
however gives an extra contribution to the higher order calculations.
In this case, therefore, to the calculation of $\Sigma^{(5)}(0)$ from the
diagrams of Fig.~2 must be added a piece from $\Sigma^{(3)}(0)$.  Here the
relevant parameters are $\mpi$ and the pseudo-vector $\pi N$ coupling
$f_{\pi\N\N}\equiv  g_{\pi\N\N}/\mn=g/F(1+O(\mpi^2))$.

\begin{figure}
  \begin{center} \mbox{\epsfig{file=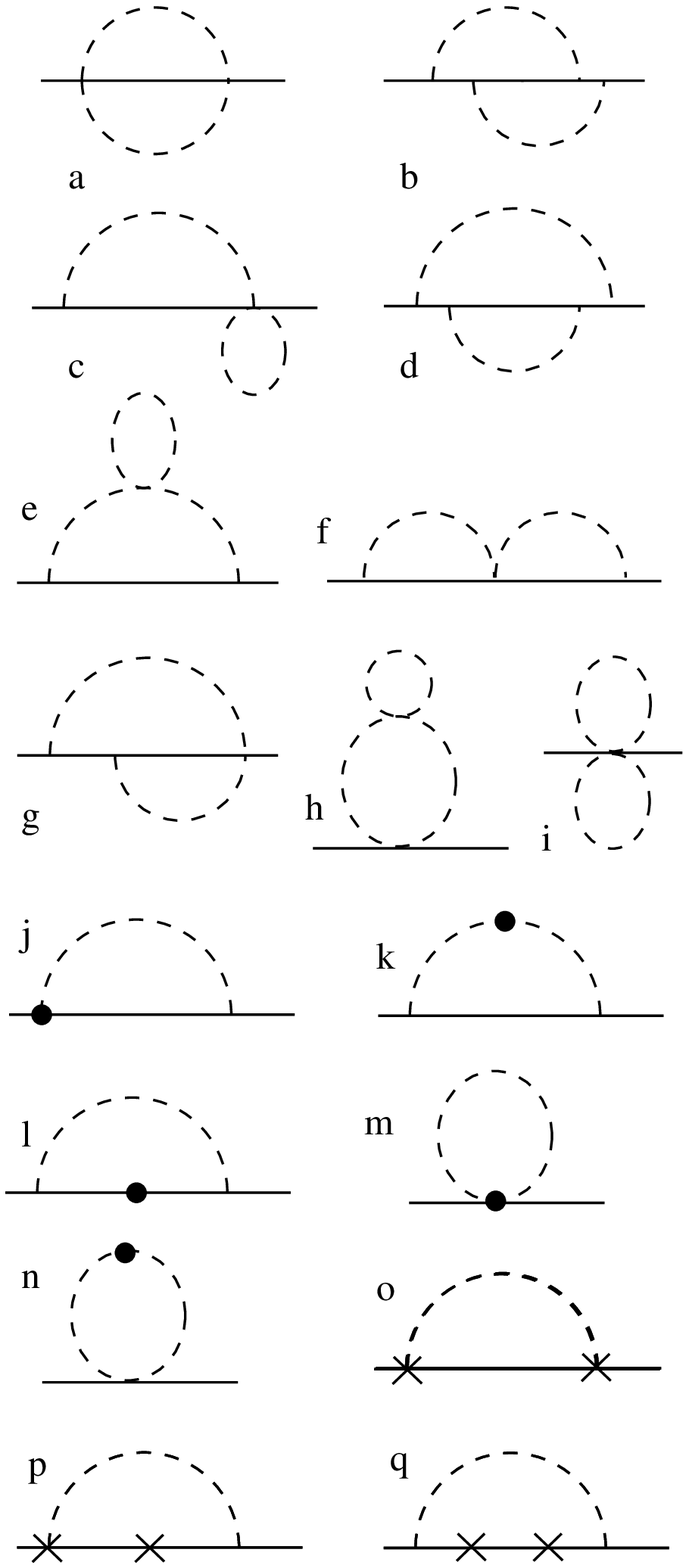,width=8.5truecm,angle=0}}
  \end{center}
{\bf Fig.~2:}                                      
Contributions to $\Sigma^{(5)}$.  Solid dots represent insertions from
${\cal L}_{\pi\N}^{(3)}$ and ${\cal L}_{\pi\pi}^{(4)}$, and crosses from ${\cal
L}_{\pi\N}^{(2)}$ (both include fixed terms from the expansion in $1/\mn$).
\end{figure}

The diagrams which contribute to  $\Sigma^{(5)}(0)$ are shown in Fig.~2.  They
consist of two-loop diagrams, and one-loop diagrams with either one insertion 
from ${\cal L}_{\pi\N}^{(3)}$ or ${\cal L}_{\pi\pi}^{(4)}$, or two
insertions from ${\cal L}_{\pi\N}^{(2)}$. In all cases only zero
external momentum is required.
The two-loop diagrams shown in Fig.~2f-i
all vanish trivially.  The diagram in Fig.~2a can after some algebra be
written as $(3M^2/4(2d-3)F^4)I$, where 
\beq
I=\int {d^d l\, d^d k \over (2\pi)^{2d}}\;{1\over v\cdot k\,(M^2-l^2) 
(M^2-(k-l)^2)}.
\eeq 
This integral can be done by using Feynman parameters, first combining the 
mesonic propagators in the standard way and integrating over $l$, 
and then using the identity
$${1\over P^p Q^q}={2^q\Gamma(p+q)\over\Gamma(p)\Gamma(q)}
\int_{0}^{\infty}{y^{q-1}dy\over(P+2y Q)^{p+q}}$$
to include the heavy-baryon propagator. Integration over $y$ and $k$
yields
\begin{eqnarray}
I&=&-{M^{2d-5}\pi^\shalf\Gamma(\frac 5 2 -d)\over (4\pi)^d}\int_0^1
(x-x^2)^{(1-d)/2}dx \nonumber\\
&=& -{M^{2d-5}\,2^{d-2}\pi\Gamma(\frac 5 2 -d)\Gamma(\frac {3-d} 2)
\over (4\pi)^d \Gamma(2-\frac d 2)}
\end{eqnarray}
which tends to zero as $d\to4$.
 
The same integral also appears in the evaluation of Fig.~2b and 2d, and it is
the only non-separable two-loop integral which does.  Since it vanishes, all
the two-loop diagrams 2a-e are proportional to $\Delta_\pi J_0(0)$, in the
notation of \cite{mei95}. The integral  $J_0(0)=-M/8\pi$ is the one which
enters the one-loop diagram, Fig.~1, and $\Delta_\pi=2M^2L(M)$ is just the
integral of the meson propagator and diverges as $1/(d-4)$.  These divergences
are cancelled by the graphs of Fig.~2j-l with counterterm insertions from 
${\cal L}_{\pi\N}^{(3)}$ and ${\cal L}_{\pi\pi}^{(4)}$, which however bring 
in the
low energy constants  $2 \overline{d}_{16} - \overline{d}_{18}$ from the
baryonic, and $\overline{l}_3$ and $\overline{l}_4$ from the mesonic
Lagrangians.  (The notation is that of \cite{mei98}, with LEC's defined
to absorb the usual factors of $\log(M/\mu)$;  in \cite{moj98}
$\overline{d}_{n} \to \overline{d}_{n+1}$ and $\overline{l}_n \to
\overline{l}_n/16\pi^2$.) The graphs with two insertions from 
${\cal L}_{\pi\N}^{(2)}$, Fig.~2o-q, are all finite. (Counterterm graphs 2m-o 
all give vanishing contributions.) The final contribution of
Fig.~2, with the conventions of \cite{mei98}, is
\begin{eqnarray}
\Sigma^{(5)}_{\rm 2-loop+CT}(0)& = &  
{3g^2M^5\over 32\pi F^2}\left({2\overline{l}_4-3\overline{l}_3\over F^2}-
{4(2 \overline{d}_{16} - \overline{d}_{18})\over g} \right.\nonumber\\&&\left.
+{g^2\over 32\pi^2F^2}+{1\over 8m^2}+{6c_1\over m}+24c_1^2\right).
\label{sig1}
\end{eqnarray}

\begin{figure}
  \begin{center} \mbox{\epsfig{file=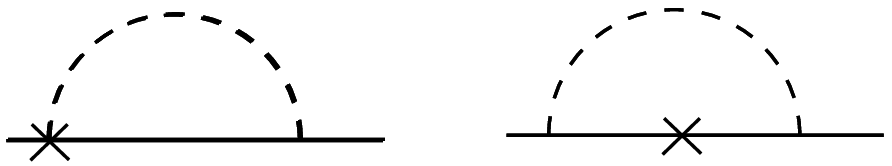,width=8.5truecm,angle=0}}
  \end{center}
{\bf Fig.~3:} Contributions to $\Sigma^{(4)}$.
\end{figure}

The next contribution to $\delta m^{(5)}$ is from the last three terms of
Eq.~\ref{mass5};  the relevant graphs are shown in Fig.~3 and Fig.~1, and the
integrals are finite.   The result is
\begin{eqnarray}
&&\delta m^{(3)}{\Sigma^{(3)}}'(0)+\delta m^{(2)}{\Sigma^{(4)}}'(0)
+\half(\delta m^{(2)})^2{\Sigma^{(3)}}''(0)\nonumber\\ &&\,\,={3g^2M^5\over 32\pi F^2} 
\left({3g^2\over32\pi^2F^2}-4c_1 \left({3\over 2m} +12c_1\right)+ 
24c_1^2\right). 
\label{sig2}
\end{eqnarray}
It may be seen that all terms involving the LEC $c_1$ vanish from the sum of 
Eqs~(\ref{sig1},\ref{sig2}).

Finally we need the contribution to $\Sigma^{(5)}(0)$ from replacing the bare
constants in $\Sigma^{(3)}(0)$ with their physical values. (The expression for 
$\Sigma^{(3)}(0)$ may be found from that for $\delta m^{(3)}$ in
Eq.~\ref{mass23} by reinstating the bare coupling constants.)  The diagrams
which contribute to the renormalised $\pi N$ coupling $f_{\pi\N\N}
(=g_{\pi\N\N}/\mn)$ are given in Fig.~4. 

The last diagram of Fig.~4 indicates the contribution from the expansion of 
the pion and nucleon wavefunction renormalisation to order $\mpi^2$. Some
comment is required about $Z_N$, which is 
defined as the residue of the nucleon propagator at the pole.  This has
recently been the subject of a paper by Ecker and \Mojzis \cite{eck97}, who
point out a correction which is necessary in order to reproduce the results of 
a relativistic calculation in HBCPT. It arises because a nucleon can also be
created by the eliminated ``small"  component of the relativistic nucleon
field, and so the normalisations of the relativistic and heavy baryons do not 
match. The same correction was included through the spinor normalisation by
Fearing {\it et al.\ }\cite{fear}.  The net effect of including the
small-component sources, to order $q^3$, is to  give
\beq
Z_\N=(1+\kvec^2/4\mn^2)Z_\N^{\rm HB},
\eeq
where $Z_\N^{\rm HB}$ is calculated purely from the HBCPT Lagrangian. 
(Whereas in the relativistic theory $Z_\N$ is a constant, in HBCPT it may
depend on the on-shell three-momentum $\kvec$.) In the framework of
Ref.~\cite{mei98}, which we have been using here, 
\beq
Z_\N^{\rm HB}=1-\kvec^2/4\mn^2 +\ldots,
\eeq
where other terms of order $q^2$ have been
suppressed.  Thus in this framework, the dependence on $\kvec$ cancels to this
order.\footnote{In \cite{eck97}, $Z_\N$ in this framework is given wrongly,
since the $O(1/m^2)$ kinetic energy insertion, Eq.~C1 of \cite{mei98}, has been
missed.  We have repeated our calculations in Ecker and \Mojzis's framework, 
and  obtained the same final result.}  With this small-component-source
correction, we find that there are no $O(1/m^2)$ corrections to $f_{\pi\N\N}$.  
(In \cite{eck97} such terms are also shown to be absent from $g_\A$,  so the
usual expression for the Goldberger-Treiman discrepancy\cite{mei95,mei98},
proportional only to $\overline{d}_{16}$, holds.)  Thus we obtain fror the
physical pion mass and $\pi N$ coupling constant,                                         
\begin{eqnarray}
\mpi^2&=&M^2(1+2\overline{l}_3M^2/F^2)\\
f_{\pi\N\N}&=&{g\over F}\left[1-M^2\left({g^2\over 16\pi^2F^2}+
{\overline{l}_4\over F^2} -{4 \overline{d}_{16} - 2\overline{d}_{18}\over
g}\right)\right]\nonumber
\end{eqnarray}
and substituting in $\sigma^{(3)}(0)$ to obtain the final contribution to
$\delta m^{(5)}$ gives 
\begin{eqnarray}
\Sigma^{(5)}_{\rm 1-loop}(0)= {3g^2M^5\over 32\pi F^2}&&\left( 
-{g^2\over 8\pi^2F^2}-{2\overline{l}_4-3\overline{l}_3\over F^2} \right.
\nonumber\\&&\quad\left. +{4(2 \overline{d}_{16} - \overline{d}_{18})\over g}
\right).
\label{sig3}
\end{eqnarray}

\begin{figure}
  \begin{center}
\mbox{\epsfig{file=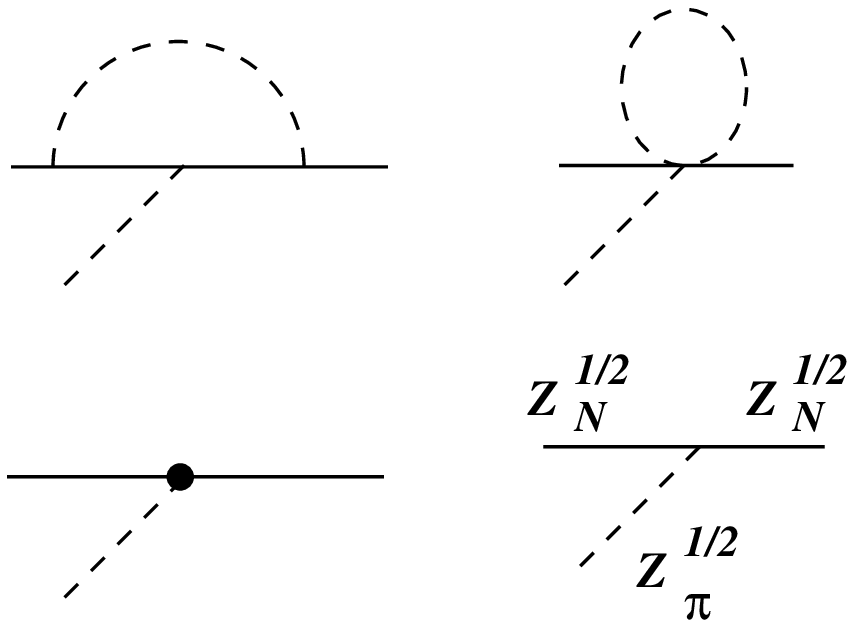,width=8.0truecm,angle=0}}
  \end{center}
{\bf Fig.~4:} Contributions to the pion-nucleon vertex of order $\mpi^2$. 
\end{figure}

Collecting all contributions, Eq.~(\ref{sig1},\ref{sig2},\ref{sig3}), we obtain
our final result,
\beq 
\delta m^{(3)}+\delta m^{(5)}=-{3g_{\pi\N\N}^2\mpi^3\over 32 \pi \mn^2}
\left(1-{\mpi^2 \over 8 \mn^2}\right).
\eeq
Thus in the heavy-baryon limit the order $\mpi^5$ contribution vanishes, with
all corrections being absorbed in the physical pion mass and pion-nucleon
coupling constant in the $\mpi^3$ contribution.  For finite nucleon mass, 
the correction is just that obtained obtained if the relativistic one-loop
contribution is expanded in powers of $\mpi/\mn$ \cite{GSS}.  (Since the
$1/\mn$ terms in the HBCPT Lagrangian are constructed to respect Lorentz
invariance, this agreement is reassuring but certainly not surprising.)

There remains the question of why there is no other fifth order correction,
apart from the renormalisation of the coupling constants of the lowest-order
theory and  $1/\mn$ corrections.  In fact the only thing which could give such
corrections would be irreducible two-loop integrals from the diagrams of
Fig.~2a, 2b and 2d.  As detailed above, only one such integral enters, and in
four dimensions it vanishes.  This vanishing seems to be essentially
accidental.  It does not occur in odd dimensions, nor in all probability for
unequal meson masses. (We have been able to calculate it 
for one massless and one massive meson in four dimensions, with a finite but
non-zero result.)

Interestingly, it has been conjectured that the main contributions at order
$q^5$ would come from insertions of vertices from ${\cal L}_{\N\pi}^{2,3}$ in 
one-loop diagrams and not from genuine two-loop diagrams \cite{bor}.  Since the
two-loop integral vanishes, this is trivially satisfied; however the extent to
which other terms are absorbed in the physical coupling constants has not
hitherto been realised.

As set out in Eq.~13, the third and fifth order pieces in the expansion of the
nucleon mass in  powers of $\mpi$ are now known.   For a full picture of the
convergence, we need to know the second and fourth order pieces.  The latter
have now been calculated,\cite{stein98} and fall into two categories; pieces
involving new  LEC's from the fourth order Lagrangian, and fixed pieces. The 
latter are very small, but the LEC's in the former are not known.  As well as
entering the nucleon mass, these affect the pion-nucleon sigma term and $\pi N$
scattering.  If $c_1$ were well determined from scattering, the analysis of
which has currently been carried out only to order $q^3$, then comparison with
the sigma term would tell us whether or not there was still a significant
discrepency to be made up from fourth or higher order contributions.
Unfortunately the analyses are not sensitive to $c_1$, though the preferred
value would leave a substantial $\mpi^4$ contribution. None-the-less it would
appear at least that the odd and even power series are converging separately.

\def\bkf{\hfill\break}

\end{document}